\documentclass[letter,bibyear]{aa} 

\usepackage{natbib}
\usepackage{graphicx}
\usepackage{txfonts}

\newcommand{\jmst}{J.~Mol.~Struct.}   

\newcommand{\kms}{km s$^{-1}$}

\begin{document}

\title{Discovery of fulvenallene in TMC-1 with the QUIJOTE$^1$ line survey
\thanks{Based on observations carried out
with the Yebes 40m telescope (projects 19A003,
20A014, 20D023, and 21A011). The 40m
radiotelescope at Yebes Observatory is operated by the Spanish Geographic 
Institute
(IGN, Ministerio de Transportes, Movilidad y Agenda Urbana).}}

\author{
J.~Cernicharo\inst{1},
R.~Fuentetaja\inst{1},
M.~Ag\'undez\inst{1},
R.~I.~Kaiser\inst{2},
C.~Cabezas\inst{1},
N.~Marcelino\inst{3,4},
B.~Tercero\inst{3,4},
J.~R.~Pardo\inst{1}, and
P.~de~Vicente\inst{3}
}

\institute{Grupo de Astrof\'isica Molecular, Instituto de F\'isica Fundamental (IFF-CSIC),
C/ Serrano 121, 28006 Madrid, Spain\\ \email jose.cernicharo@csic.es
\and Department of Chemistry, University of Hawaii at Manoa, Honolulu, HI 96822, USA
\and Observatorio de Yebes, IGN, Cerro de la Palera s/n, 19141 Yebes, Guadalajara, Spain
\and Observatorio Astron\'omico Nacional, IGN, C/Alfonso XII 3, 28014 Madrid, Spain
}

\date{Received; accepted}

\abstract{
We report the detection of fulvenallene ($c$-C$_5$H$_4$CCH$_2$) in the direction of TMC-1
with the QUIJOTE line survey. Thirty rotational transitions with $K_a$=0,1,2,3 and $J$=9-15
were detected. The best rotational temperature 
fitting of the data is 9\,K and a derived column density is (2.7$\pm$0.3)$\times$10$^{12}$ cm$^{-2}$, which is only a factor of  4.4 below that of its potential precursor cyclopentadiene ($c$-C$_5$H$_6$), and 
1.4--1.9 times higher than that of the ethynyl derivatives of cyclopentadiene. We searched for
fulvene ($c$-C$_5$H$_4$CH$_2$), a CH$_2$ derivative of cyclopentadiene, for which we derive a 3$\sigma$ 
upper limit to its
column density of (3.5$\pm$0.5)$\times$10$^{12}$ cm$^{-2}$.
Upper limits were also obtained for toluene (C$_6$H$_5$CH$_3$) and styrene (C$_6$H$_5$C$_2$H$_3$), 
the methyl and vinyl derivatives of benzene. Fulvenallene  and  ethynyl cyclopentadiene 
are likely formed in the reaction between cyclopentadiene ($c$-C$_5$H$_6$) and the ehtynyl radical
(CCH). However, the bottom-up gas-phase 
synthesis of cycles in TMC-1 underestimates the abundance of cyclopentadiene by two orders of 
magnitude, which strengthens the need to study all possible chemical pathways to cyclisation in 
cold dark cloud environments, such as TMC-1. However, the inclusion of the reaction between C$_3$H$_3^+$ and C$_2$H$_4$
produces a good agreement between model and observed abundances.
}


\titlerunning{Fulvenallene in TMC-1}
\authorrunning{Cernicharo et al.}

\maketitle

\section{Introduction}

The QUIJOTE\footnote{\textbf{Q}-band \textbf{U}ltrasensitive \textbf{I}nspection \textbf{J}ourney 
to the \textbf{O}bscure \textbf{T}MC-1 \textbf{E}nvironment} 
line survey of \mbox{TMC-1} \citep{Cernicharo2021a} performed with the Yebes 40m radio telescope has
permitted the  detection of  nearly 40 new molecular species in recent months, most of which are
hydrocarbons and cycles such as
indene, benzyne, cyclopentadiene, and two isomers of ethynyl cyclopentadiene (see e.g. 
\citealt{Cernicharo2021a,Cernicharo2021b,Cernicharo2021c,Cernicharo2021d,Cernicharo2021e} 
and references therein). Propargyl (CH$_2$CCH) has been found to be one of the most abundant hydrocarbon radicals 
in this source
\citep{Agundez2021,Agundez2022}. Other hydrocarbons
such as ethynylallene (CH$_2$CCHCCH), vinylacetylene (CH$_2$CHCCH), and butadiynylallene (CH$_2$CCHC$_4$H) 
have also been found to have   high abundances \citep{Cernicharo2021b,Cernicharo2021d,
Fuentetaja2022}. Moreover, cyano derivatives of benzene and naphthalene have also been found towards TMC-1
\citep{McGuire2018,McGuire2021}. This suggests that benzene and naphthalene are also very abundant in this cold
prestellar core. To understand 
the formation of these species in the gas-phase, and their large abundances (see Appendix \ref{abundance_hydrocarbons}), 
we have to consider very fast reactions between a few precursors.
It is possible, however,  that some gas-phase chemical reactions are still missing in the chemical networks. It is 
also likely that
important hydrocarbons with low dipole moments are still awaiting detection. Hence,
discovering new hydrocarbons through sensitive line surveys will certainly help in getting a complete chemical 
network (gas-phase and dust surface reactions) to explain the chemistry of hydrocarbons and polycyclic aromatic hydrocarbons (PAHs) in TMC-1.

\begin{figure}
\centering
\includegraphics[width=0.2\textwidth]{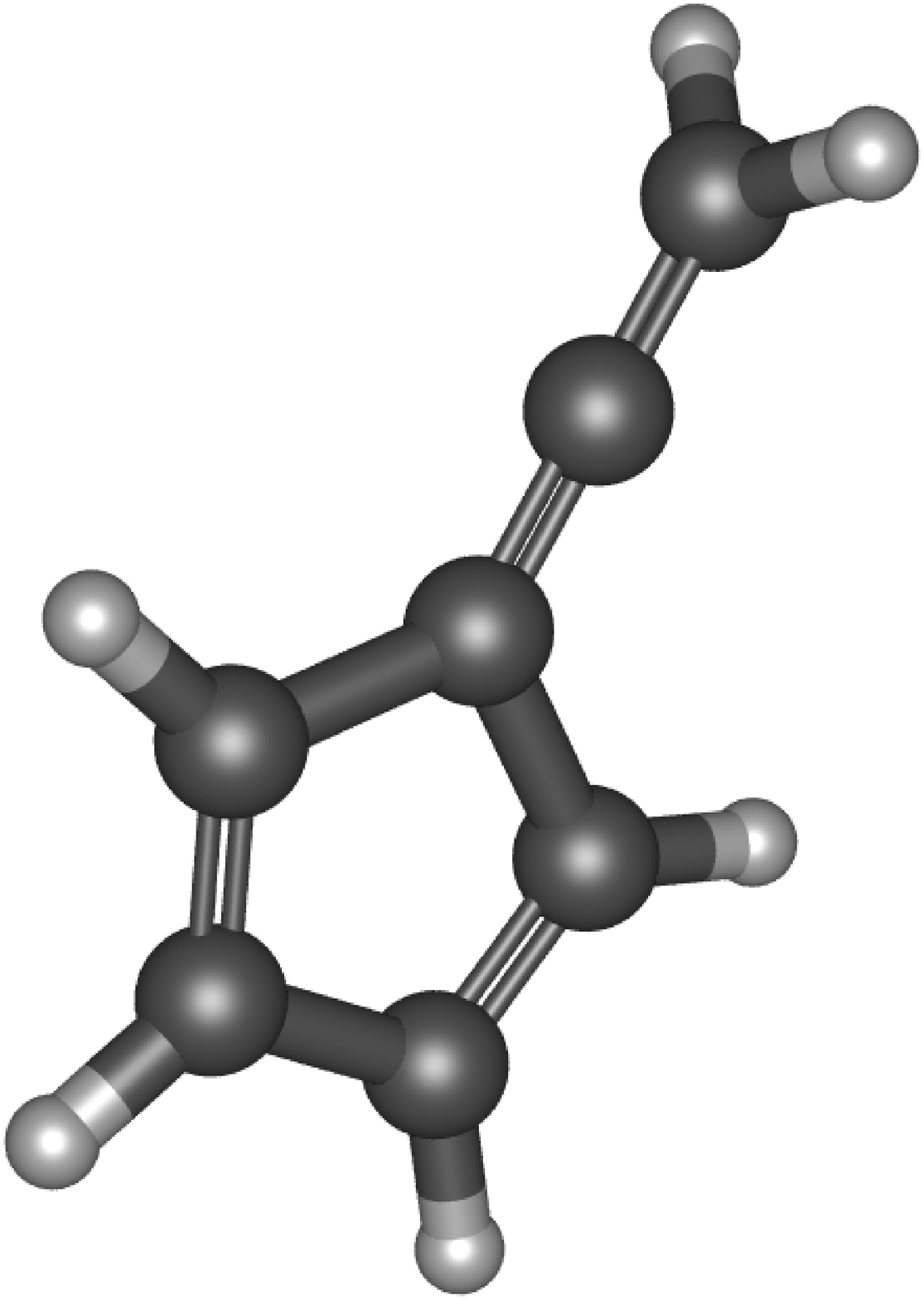}
\caption{Structure of fulvenallene ($c$-C$_5$H$_4$CCH$_2$).}
\label{fig_structures}
\end{figure}

In this letter we report the discovery of fulvenallene ($c$-C$_5$H$_4$CCH$_2$) through the detection of 30
rotational transitions in the frequency range 31--50 GHz. The structure of fulvenallene ($c$-C$_5$H$_4$CCH$_2$), 
an isomer of ethynyl cyclopentadiene, is shown in Fig.~\ref{fig_structures}. We discuss the chemistry of this 
hydrocarbon cycle in the context of a bottom-up gas-phase scheme for the formation of hydrocarbon cycles in TMC-1.

\section{Observations} \label{observations}

New receivers built within the Nanocosmos project\footnote{\texttt{https://nanocosmos.iff.csic.es/}},
and installed at the Yebes 40m radiotelescope, were used
for the observations of \mbox{TMC-1}
($\alpha_{J2000}=4^{\rm h} 41^{\rm  m} 41.9^{\rm s}$ and $\delta_{J2000}=
+25^\circ 41' 27.0''$). A detailed description of the system is 
given by \citet{Tercero2021}. Details of the QUIJOTE line survey are provided by \citet{Cernicharo2021a}.
The observations were carried out during different observing runs between November 2019 and May 2022.
The receiver consists of two cold high electron mobility transistor amplifiers covering the
31.0--50.3 GHz band with horizontal and vertical             
polarisations. Receiver temperatures in the runs achieved during 2020 vary from 22 K at 32 GHz
to 42 K at 50 GHz. Some power adaptation in the down-conversion chains  reduced
the receiver temperatures during 2021 to 16\,K at 32 GHz and 25\,K at 50 GHz.
The backends are $2\times8\times2.5$ GHz fast Fourier transform spectrometers
with a spectral resolution of 38.15 kHz
providing the whole coverage of the Q band in each polarisation.

The data presented here
correspond to 546 hours of observing time on the source, of which 293 and 253 hours were acquired with a 
frequency switching throw of 8 MHz and 10 MHz, respectively. The data analysis procedure is described
in Appendix \ref{data_analysis}.  
The intensity scale used in this work, antenna temperature
($T_A^*$), was calibrated using two absorbers at different temperatures and the
atmospheric transmission model ATM \citep{Cernicharo1985, Pardo2001}.
The antenna temperature has an estimated uncertainty of 10~\% and can be 
converted to main beam brightness temperature, $T_{mb}$, by dividing by $B_{\rm eff}$/$F_{\rm eff}$. For the Yebes 40m 
telescope, $B_{\rm eff}$\,=\,0.738\,$\exp$[$-$($\nu$(GHz)/72.2)$^2$] and $F_{\rm eff}$\,=\,0.97 \citep{Tercero2021}.
All data were analysed using the GILDAS package\footnote{\texttt{http://www.iram.fr/IRAMFR/GILDAS}}.

\begin{figure}
\centering
\includegraphics[width=0.49\textwidth]{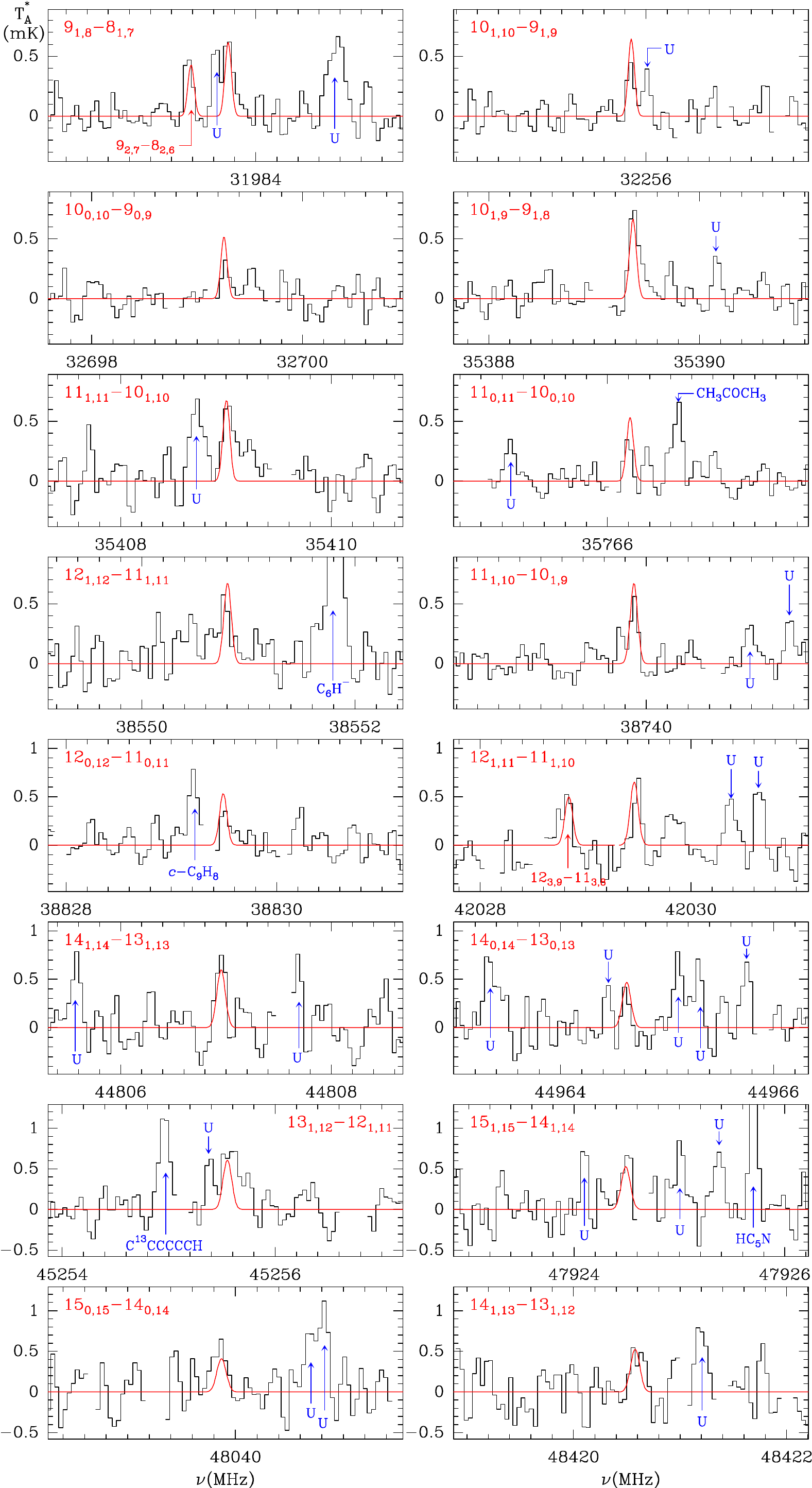}
\caption{Observed lines with $K_a$=0,1 of fulvenallene($c$-C$_5$H$_4$CCH$_2$) towards \mbox{TMC-1} (see line parameters  in Table \ref{line_par_fva}).
The abscissa corresponds to the rest frequency assuming a local standard of rest velocity of 5.83
km s$^{-1}$. 
The ordinate is the antenna temperature corrected for atmospheric and telescope losses in mK.
The red line shows the synthetic spectrum derived for
T$_{rot}$=10\,K and N($c$-C$_5$H$_4$CCH$_2$)=2.7$\times$10$^{12}$ cm$^{-2}$.
Blank channels correspond to negative features produced
in the folding of the frequency switching data. Some $K_a$=2,3 lines also appear in these panels
(top left and fourth  from bottom right). Additional $K_a$=2,3 lines are shown in Fig. \ref{fig_fva_2}.}
\label{fig_fva_1}
\end{figure}

\begin{figure}
\centering
\includegraphics[width=0.49\textwidth]{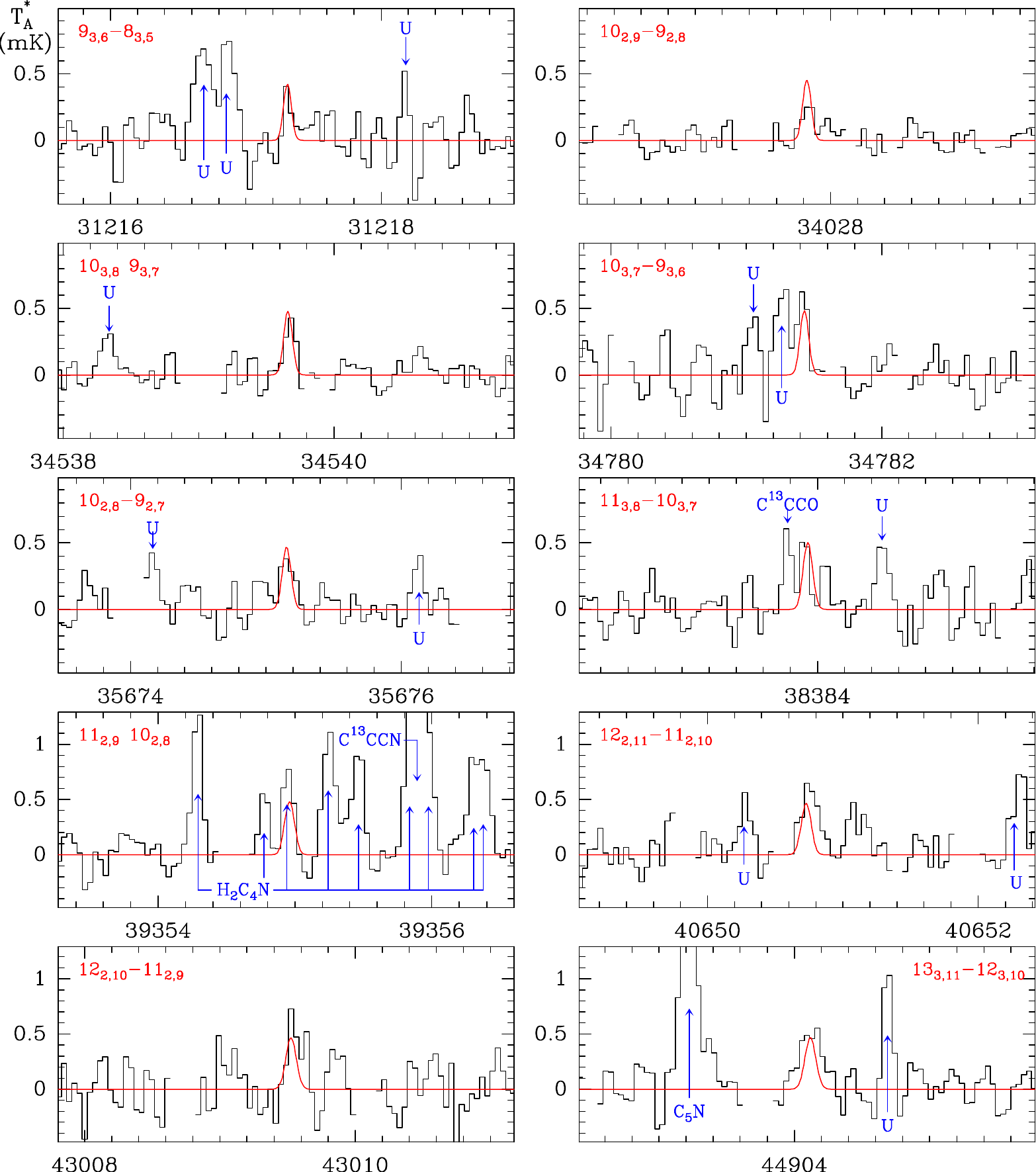}
\caption{Same as in Figure 1, but for lines  $K_a$=2 and 3 (see line parameters  in Table \ref{line_par_fva}).
}
\label{fig_fva_2}
\end{figure}

\section{Results} \label{results}
QUIJOTE has now reached  a level of sensitivity (0.1--0.3 mK per 38.15 kHz channel across the Q band) that permits 
the  detection of  new
isotopologues and derivatives of abundant species. Examples of detection and complete spectral characterisation 
of rare isotopologues through QUIJOTE data are
HDCCN \citep{Cabezas2021a}, CH$_2$DC$_3$N \citep{Cabezas2021b}, and CH$_2$DC$_4$H \citep{Cabezas2022a}. 
Although QUIJOTE has not yet reached the confusion limit, special care should  be taken
when assigning lines to a given molecule as blending with other features often
occurs. Line identification in this work was done using the following catalogues: 
MADEX \citep{Cernicharo2012}, CDMS \citep{Muller2005}, and JPL \citep{Pickett1998}. 
By May 2022 the MADEX catalogue had 6434 spectral
entries corresponding to the ground state and vibrationally excited state, together
with the corresponding isotopologues, of 1734 molecules. 

Rotational spectroscopy of fulvenallene was   performed in the laboratory by \citet{Sakaizumi1993} up 
to frequencies
of 39.3 GHz with an accuracy of 50 kHz, and by \citet{McCarthy2020} up to frequencies of 25 GHz with an 
accuracy of 2 kHz.
To search for fulvanellene in TMC-1 we fitted these data, but discarded the less accurate measurements
of \citet{Sakaizumi1993}, which are also present in the data set of \citet{McCarthy2020}. The derived 
rotational and distortion
constants are shown in the middle column of Table \ref{line_fit_fva}, and were implemented in the MADEX code 
to predict the spectrum. Fulvenallene has $C_{2v}$ symmetry, hence we have included the statistical
weights for the ortho and para species, 9 for ortho ($K_a$ odd) and 7 for para ($K_a$ even). The energy
difference between the two states is 0.47\,K.

We first searched for the strongest lines, which correspond to transitions with $K_a$=0 and 1. 
All $K_a$=0,1 lines within the Q band corresponding to transitions with $J$ between 9 and 14 (a total of sixteen lines) 
are detected.
These lines were found within 10--20 kHz of the predicted frequencies and are shown in Fig. \ref{fig_fva_1}. Their
line parameters are given in Table \ref{line_par_fva}. Once the $K_a$=0 and 1 lines were detected we searched for the 
less intense $K_a$=2 and 3 transitions.
Fourteen of these lines were detected and are shown in Fig. \ref{fig_fva_2};  their line
parameters are given in Table \ref{line_par_fva}. We note that two of these lines are shown in Fig. \ref{fig_fva_1} in the upper left panel and the fourth  panel 
from the  bottom on the right.
Although the differences between the observed and predicted frequencies are small, they show a systematic behaviour; 
in particular, some
of the $K_a$=2,3 lines show discrepancies
with respect to their frequencies as large as 50 kHz. Taking into account that the predictions
in the Q band are strongly dependent on the low accuracy data of \citet{Sakaizumi1993} we fitted the frequencies of the
TMC-1 lines alone to a Watson Hamiltonian in reduction A representation I$^r$ \citep{Watson1977} using the FITWAT 
code \citep{Cernicharo2018}. 
This code reproduces the same results as the standard SPFIT code of \citet{Pickett1991}, as detailed in  Appendix A
of \citet{Cernicharo2018}. The frequency measurements have been weighted 
as 1/$\sigma^2$, where $\sigma$ is the estimated frequency uncertainty.
The resulting parameters are given in the first column
of Table \ref{line_fit_fva}. The comparison with the fit to the laboratory data (middle column of the same Table) 
indicates
that all the observed lines can be reproduced with rotational and distortion constants identical to those of the
laboratory within the derived uncertainties of the different constants. Hence, the detection of fulvenallene is solid 
as it is based on 30 individual lines. Finally, taking into account that the accuracy of our measurements in the 
31--50 GHz
domain is better than those of \citet{Sakaizumi1993}, we performed a merged fit to the laboratory and space frequencies
of fulvenallene. The resulting constants are given in the third column of Table \ref{line_fit_fva} and represent a
significant improvement with respect to the constants obtained from the laboratory data alone (middle column of Table 
\ref{line_fit_fva}).
The constants resulting from this merged fit are the ones we recommend to predict the spectrum of fulvenallene. 
The observed and calculated frequencies of the lines included in the fit, and the difference
between the observed and calculated frequencies are given in Table \ref{line_fit_dif_fva}.

\begin{table}
\caption{Derived rotational constants for $c$-C$_5$H$_4$CCH$_2$}
\label{line_fit_fva}
\small
\centering
\begin{tabular}{{l@{\hspace{0.2cm}}ccc}}
\hline
Constant$^a$                  &      TMC-1$^b$      & Laboratory$^c$          & Merged Fit$^d$\\
       (MHz)                  &                                                                \\
\hline
$A$                           & 8180.862(38)       &    8180.813(24)       &   8180.855(20)   \\
$B$                           & 1886.66733(64)     &    1886.67218(25)     &   1886.67152(19) \\
$C$                           & 1549.11498(74)     &    1549.11344(26)     &   1549.11379(17) \\
$\Delta_J$ $\times$ 10$^5$    &    3.61(20)        &       4.46(15)        &      4.263(70)   \\
$\Delta_{JK}$ $\times$ 10$^3$ &    3.007(70)       &       3.0085(70)      &      3.0065(77)  \\
$\delta_J$   $\times$ 10$^5$  &                    &       1.07(19)        &      0.602(60)   \\
\hline
\hline
N$_{lines}$$^e$               &      30             &      97              &     127          \\
$rms^f$                 &      19             &      34              &      31          \\
$J_{max}$$^g$           &      15           &      11           &    15         \\
$K_{a, max}$$^g$           &      3           &      10           &    10         \\
$\nu_{max}$$^h$         & 48420.601           &     39354.900        &    48420.601     \\   
                   \hline
\end{tabular}
\tablefoot{\\
        \tablefoottext{a}{Fitted rotational and distortion constants. 
        The values in brackets correspond to the uncertainties of the parameters in units of the last significant digits.}
        \tablefoottext{b}{Fit to the line frequencies observed in TMC-1 alone.} 
        \tablefoottext{c}{Fit to the laboratory data (see text).}
        \tablefoottext{d}{Merged fit to the laboratory and TMC-1 data (see Table \ref{line_par_fva}).}
        \tablefoottext{e}{Number of lines in the fit.}
        \tablefoottext{f}{The standard deviation of the fit in kHz.}
        \tablefoottext{g}{Maximum value of $J$ and $K_a$ of the lines included in the fit.}
        \tablefoottext{h}{Maximum frequency of the lines included in the fit in MHz.}
}
\end{table}

Once we are confident of the detection of fulvenallene in TMC-1, we fitted the observed lines 
using a line profile
fit method \citep{Cernicharo2021e}, which permits    the rotational temperature and
the column density to be fit, assuming that  all levels have the same rotational temperature. 
A value of the dipole moment $\mu_a$=0.69$\pm$0.10\,D was used \citep{Sakaizumi1993}.
We obtained a rotational temperature of 9$\pm$1\,K and a column density
of (2.7$\pm$0.3)$\times$10$^{12}$ cm$^{-2}$. A rotational temperature lower than   8\,K cannot 
reproduce the observed
intensities of the $K_a$=2,3 lines.

We also searched for fulvene ($c$-C$_5$H$_4$CH$_2$), for which the rotational spectroscopy has been 
accurately recorded by \citet{McCarthy2020}. The
dipole moment of this molecule is $\mu_a$=0.424$\pm$0.001\,D \citep{Baron1972}, and hence for a column 
density similar to that of fulvenallene the lines
will be 2.6 times weaker. Although some K$_a$=0,1 lines are detected at a 3$\sigma$ level we conclude 
that the molecule is below the
present sensitivity limit of QUIJOTE. 
We obtain a 3$\sigma$ upper limit to the column density of 3.5$\times$10$^{12}$ cm$^{-2}$ for fulvene.

Two derivatives of benzene are also of interest in the context of this work, toluene ($c$-C$_6$H$_5$CH$_3$) and 
styrene ($c$-C$_6$H$_5$C$_2$H$_3$).
For toluene, we used the laboratory data of \citet{Kisiel2004} and the dipole moment obtained by 
\citet{Rudolph1967}, 
$\mu_a$=0.38$\pm$0.01\,D. For styrene, laboratory spectroscopy up to 18 GHz is available from 
\citet{Caminati1988}, who also
derived the $a$ and $b$ components of the dipole moment as $\mu_a$=0.122$\pm$0.001\,D and 
$\mu_b$$<$0.02\,D. Due to the low dipole moment of the two species we derived relatively high 
3$\sigma$ upper limits, 6$\times$10$^{12}$ cm$^{-2}$ for toluene 
and 10$^{14}$ cm$^{-2}$ for styrene. For both species we adopted a rotational temperature of 9\,K.

\section{Discussion}

\begin{figure*}
\centering
\includegraphics[width=0.95\textwidth]{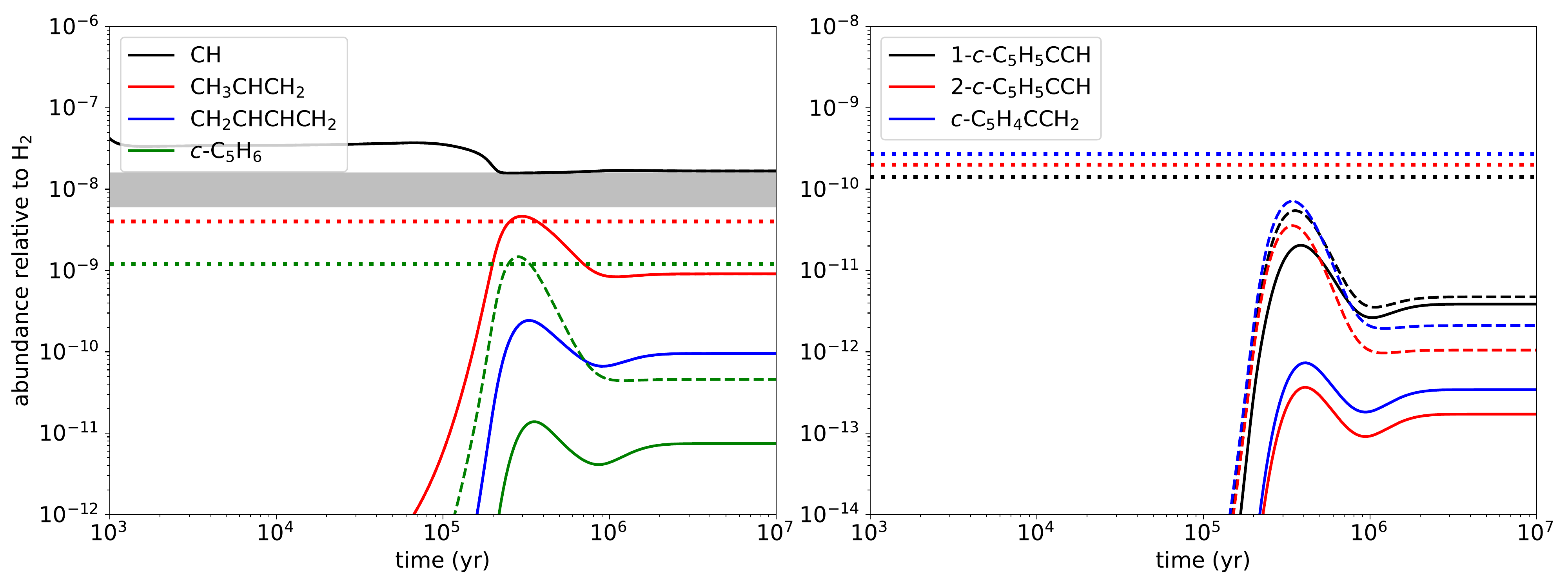}
\caption{Calculated abundances for $c$-C$_5$H$_6$ and its precursors (left panel) and for the products of its reaction with CCH (right panel). Solid lines correspond to the standard chemical model, while dashed lines correspond to a model including formation of $c$-C$_5$H$_6$ involving the reaction $l$-C$_3$H$_3^+$ + C$_2$H$_4$ (see text). The horizontal band and dotted lines correspond to the abundances observed in TMC-1 (see Table~\ref{abundances}).}
\label{fig_abun}
\end{figure*}

In the frame of a gas-phase bottom-up scenario for the formation of cycles in TMC-1, we ran 
chemical model calculations similar to those presented in \cite{Cernicharo2021f}. Here we focus 
on cyclopentadiene and derivatives. Cyclopentadiene can be formed by two successive reactions 
involving the radical CH
\begin{equation}
\rm CH_3CHCH_2 + CH \rightarrow \rm CH_2CHCHCH_2 + H,
\end{equation}
\begin{equation}
\rm CH_2CHCHCH_2 + CH \rightarrow \rm c-C_5H_6 + H,
\end{equation}
for which we adopted rate coefficients at 10\,K of 3.3\,$\times$\,10$^{-10}$ cm$^3$ s$^{-1}$ 
\citep{Loison2017} and 4\,$\times$\,10$^{-10}$ cm$^3$ s$^{-1}$ \citep{Cernicharo2021f}, respectively. 
In turn, the reaction of $c$-C$_5$H$_6$ with CCH can yield various isomers of ethynyl cyclopentadiene, 
as well as fulvenallene
\begin{subequations}
\begin{align}
\rm c-C_5H_6 + CCH & \rightarrow \rm 1-c-C_5H_5CCH + H, \\
                                   & \rightarrow \rm 2-c-C_5H_5CCH + H, \\
                                   & \rightarrow \rm c-C_5H_4CCH_2 + H.
\end{align}
\end{subequations}
We assumed a rate coefficient of 4\,$\times$\,10$^{-10}$ cm$^3$ s$^{-1}$ for reaction (3) and 
branching ratios of 0.25, 0.25, and 0.50 for channels (a), (b), and (c), respectively, where a 
higher branching ratio is adopted for fulvenallene because it is more stable than the isomers 
of ethynyl cyclopentadiene.

The synthesis of $c$-C$_5$H$_6$  therefore relies on 1,3-butadiene (CH$_2$CHCHCH$_2$; non-polar and thus invisible at 
radio wavelengths), which in turn is formed from propene (CH$_2$CHCH$_3$), whose abundance is well constrained in 
TMC-1 \citep{Marcelino2007}. In the chemical model presented in \cite{Cernicharo2021f} this pathway 
produced cyclopentadiene with an abundance in agreement with observations because propene was formed 
very efficiently due to two consecutive radiative associations of CH$_2$CCH$^+$ with H$_2$ 
\citep{Herbst2010}, followed by dissociative recombination of C$_3$H$_7^+$. This pathway is 
included in the chemical networks UMIST RATE12 \citep{McElroy2013} and {\small \texttt{kida.uva.2014}} 
\citep{Wakelam2015} and yields a very high propene abundance (see e.g. \citealt{Agundez2013}). 
However, the radiative associations mentioned above were later on shown to be inefficient at low 
temperatures \citep{Lin2013}. If they are neglected, the calculated abundance of propene drops by 
many orders of magnitude, well below the observed value, with drastic implications for the chemistry 
of many other hydrocarbons, such as 1,3-butadiene and cyclopentadiene, which are direct descendants 
of propene.

Currently, no efficient gas-phase route to propene has been identified in cold dark clouds. However, 
we know from observations that propene is present with a fairly large abundance. To overcome the lack 
of formation of propene in our gas-phase chemical model, we fixed the peak abundance of propene to 
be in agreement with the value observed in TMC-1. The results from the chemical model are shown as 
solid lines in Fig.~\ref{fig_abun}. It is seen that 1,3-butadiene is formed with an abundance ten 
times lower than propene, while $c$-C$_5$H$_6$ in turn reaches an abundance ten times below that of 
1,3-butadiene. However, according to the observations, the abundance of $c$-C$_5$H$_6$ is only slightly 
below that of propene. The chemical model thus underestimates the abundance of $c$-C$_5$H$_6$ by 
two orders of magnitude, and this translates directly to the derivatives of cyclopentadiene 
resulting from its reaction with CCH. In the case of 1-$c$-C$_5$H$_5$CCH, its calculated abundance 
is higher than that of 2-$c$-C$_5$H$_5$CCH and fuvenallene because it is   formed through reaction (3), and in addition by the reaction between C$_3$H and 
1,3-butadiene (see \citealt{Cernicharo2021e}).

The fact that 1,3-butadiene and cyclopentadiene are first- and second-generation descendants of 
propene means that they reach progressively lower abundances. The resulting abundances are 
essentially controlled by the rates of formation, through reactions (1) and (2), and the 
rates of destruction, which in the chemical model is dominated by reaction with atomic carbon. 
We assumed rate coefficients of 10$^{-10}$ cm$^3$ s$^{-1}$ for the reactions of atomic carbon 
with 1,3-butadiene and cyclopentadiene, in line with the values determined experimentally at low 
temperature for reactions between C and unsaturated hydrocarbons \citep{Chastaing2001}. In summary, 
regardless of what  the true pathway to propene is, the consecutive reactions with CH are found to be 
insufficient to form cyclopentadiene with the observed abundance.

We examined possible formation routes to $c$-C$_5$H$_6$ involving cations. Among them, the reaction 
between $l$-C$_3$H$_3^+$ and C$_2$H$_4$ has been measured to be rapid, with a rate coefficient of 
1.1\,$\times$\,10$^{-9}$ cm$^3$ s$^{-1}$, and the ion C$_5$H$_7^+$ has been observed as product 
\citep{Smyth1982,Anicich2003}. Including this reaction in the chemical model and assuming that 
the ion C$_5$H$_7^+$ is cyclic and recombines dissociatively with electrons to yield $c$-C$_5$H$_6$ 
with a standard rate coefficient of 10$^{-7}$ ($T$/300)$^{-0.5}$ cm$^3$ s$^{-1}$, the calculated peak 
abundances of cyclopentadiene, and those of its CCH derivatives $c$-C$_5$H$_5$CCH and 
$c$-C$_5$H$_4$CCH$_2$, become close to the observed values (see dashed lines in Fig.~\ref{fig_abun}). 
Although our assumptions on the $l$-C$_3$H$_3^+$ + C$_2$H$_4$ reaction require further experimental or 
theoretical support, this suggests that ion-neutral reactions can provide efficient formation routes 
to cycles in TMC-1. We note that $c$-C$_3$H$_3$$^+$ does not react with C$_2$H$_4$
\citep{Smyth1982,Anicich2003}.

\begin{figure}
\centering
\includegraphics[width=0.49\textwidth]{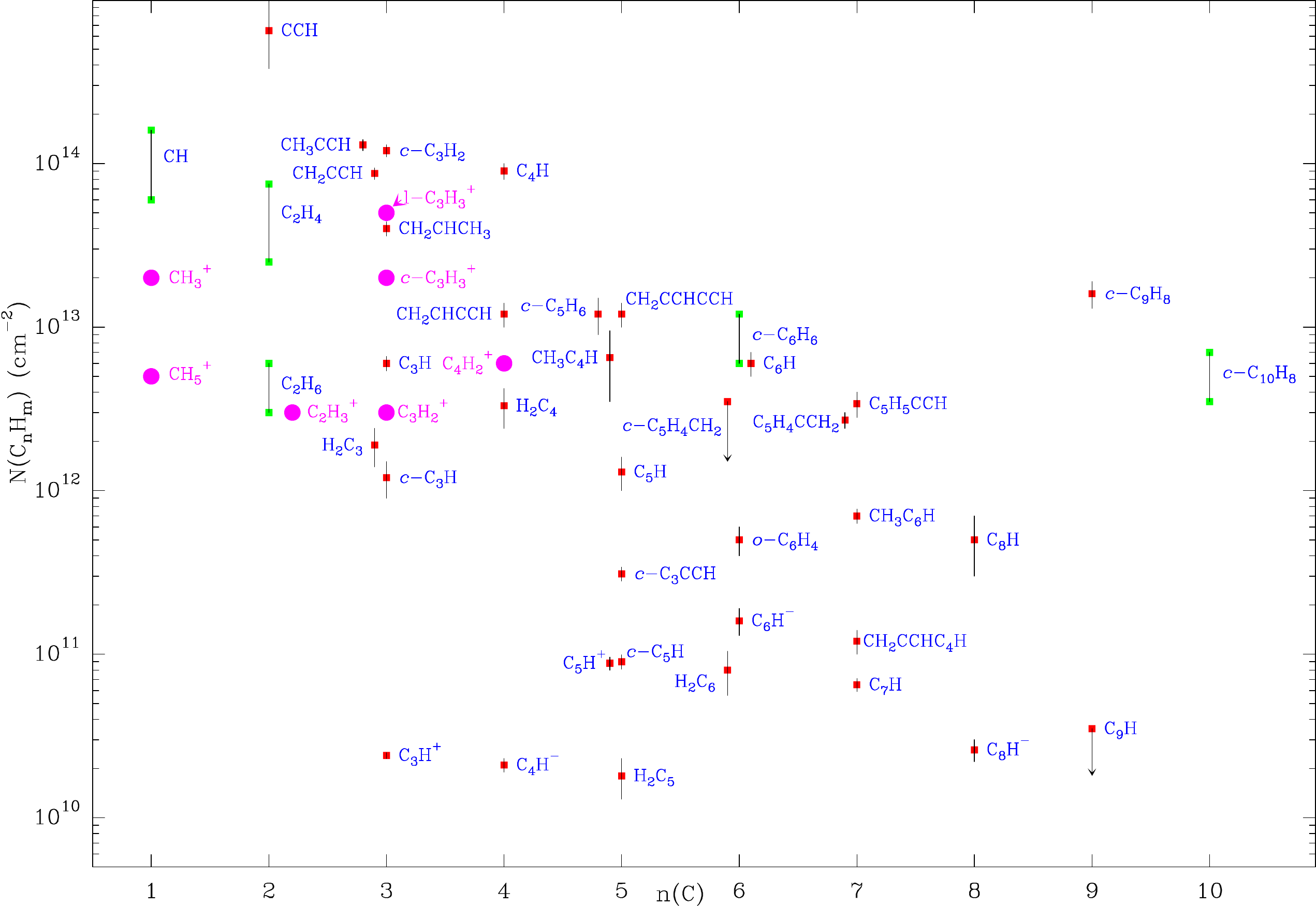}
\caption{Graphic representation of the abundances of C$_n$H$_m$ hydrocarbons
from Table \ref{abundances} in Appendix \ref{abundance_hydrocarbons}. The abscissa
corresponds to the number of carbon atoms in the molecule and the ordinate to
the column density of each hydrocarbon. For species
whose  abundance has been estimated from cyano derivatives the upper and lower
ranges are indicated by two green squares. For the other species the abundance is
indicated by a single red square and the error is represented by a vertical
black line. Abundances of the most abundant cations, according to the chemical model (see text), 
are indicated in  magenta.}
\label{fig_cnhm}
\end{figure}

To have some insight into the possible precursors of hydrocarbon cycles we collected abundances for 
hydrocarbons C$_n$H$_m$ detected in TMC-1. The abundances derived from QUIJOTE data
from previous publications of our team related to newly detected hydrocarbons, and for species
already known but with transitions in our frequency range, are given in Table \ref{abundances} and
represented in  Fig. \ref{fig_cnhm}. For species such as benzene or naphthalene we adopted
an abundance ratio between them and their cyano derivatives of 5-10. We note however that these
estimates are very uncertain. 
From Fig. \ref{fig_cnhm}
it appears that the most abundant hydrocarbon radicals are 
methylidyne (CH), ethynyl (CCH), propargyl (CH$_2$CCH), and butadiynyl (C$_4$H). The most
abundant closed shell hydrocarbons are methylacetylene (CH$_3$CCH), cyclopropenylidyne ($c$-C$_3$H$_2$), 
ethylene (C$_2$H$_4$), and propene (CH$_2$CHCH$_3$).
Indene is the most abundant cycle with more than three carbon atoms, followed by
cyclopentadiene, benzene, naphthalene, and benzyne.
If these rings are formed in the gas phase,
their precursors should be among the abundant species with 3--5 carbon atoms. 
The role of hydrocarbon cations (C$_n$H$_m$$^+$) in the chemistry of hydrocarbons has been quite 
well studied,
but their abundances remain unknown. Only C$_3$H$^+$ and C$_5$H$^+$ have been detected in TMC-1 so far
\citep{Cernicharo2022}.
Moreover, the abundance of important species for hydrocarbon chemistry, such as CH$_3$, CH$_4$, 
C$_2$H$_2$, CH$_2$CH$_2$, CH$_3$CH$_3$, and CH$_3$CH$_2$CH$_3$, remain unknown because they are 
non-polar, and thus cannot be observed at radio wavelengths. For CH$_2$CH$_2$ we used the derived 
abundance of vinyl acetylene (see Appendix
\ref{abundance_hydrocarbons}) since molecular beam experiments and electronic structure calculations 
provide compelling evidence that
it is formed in the reaction of ethylene and the ethynyl radical \citep{Zhang2009}.
Similarly, for CH$_3$CH$_3$ we can estimate its abundance from that of CH$_3$CH$_2$CCH 
\citep{Cernicharo2021d}. These estimates
are, as for benzene and naphthalene, very uncertain (see Appendix \ref{abundance_hydrocarbons}).

\section{Conclusions}

We detected fulvenallene in TMC-1, with an abundance just 4.4 times lower than its 
possible precursor cyclopentadiene and similar to those of the two ethynyl cyclopentadiene 
isomers detected in this source. Fulvenallene and  ethynyl cyclopentadiene are most 
likely formed by the reaction between cyclopentadiene and CCH. However, our chemical model, 
mainly based on neutral-neutral gas-phase reactions, undererstimates the abundance of cyclopentadiene 
by two orders of magnitude. We suggest that ion-neutral gas-phase reactions between the most abundant
cations and hydrocarbons, open and closed shell species, may provide efficient 
routes to cyclopentadiene and other cycles in TMC-1.

\begin{acknowledgements}

We thank ERC for funding through grant ERC-2013-Syg-610256-NANOCOSMOS. We also 
thank Ministerio de Ciencia e Innovaci\'on of Spain (MICIU) for funding support 
through projects PID2019-106110GB-I00, PID2019-107115GB-C21, and PID2019-106235GB-I00. 

\end{acknowledgements}

\clearpage
\onecolumn

\begin{appendix}

\section{Abundance of the C$_n$H$_m$ hydrocarbons in TMC-1} \label{abundance_hydrocarbons}

In order to have a global view of the abundances of hydrocarbons in TMC-1, we show in Table \ref{abundances} the
values obtained with QUIJOTE for most species of interest related to this work. In all cases
we assumed a source diameter of 80$''$ \citep{Fosse2001}.
Some of the species given in this table have rotational transitions within the QUIJOTE survey, but their data have not been published yet.
Most of them have been previously detected by other authors. Nevertheless, we think it  more valuable to provide column densities 
using the QUIJOTE data to have a coherent criteria in the determination of their column densities, and to use data gathered with the
same radio telescope. The data on these species will be published elsewhere. Here we comment briefly on the assumptions
adopted for these molecules to derive the column densities given in Table \ref{abundances}. 
In all cases, except when indicated, the adopted line width at half intensity, $\Delta$v, is 0.6 km s$^{-1}$.

For $l$-C$_3$H, only one rotational transition, $J$=3/2-1/2, with six fine and hyperfine components is 
covered within the QUIJOTE frequency range. 
All these lines are detected with very high  S/N values. We adopted T$_{rot}$=7 K for linear C$_3$H. 
The derived column density is (6.0$\pm$0.6)$\times$10$^{12}$ cm$^{-2}$.
We note that the
column density is not very sensitive to the adopted rotational temperature for T$_{rot}$ between 6 and 10 K. This is due to
the low energy (E$_u$=1.57 K) of the upper level $J$=3/2 (see Appendix A of \citealt{Cernicharo2021g}).

Only the rotational transition 2$_{11}$-2$_{12}$ of $c$-C$_3$H falls within the QUIJOTE survey. 
It exhibits several fine and hyperfine components. The energy or the upper level of 6.52\,K, hence
slightly more sensitive to the adopted rotational temperature than C$_3$H.
The derived column density is (1.2$\pm$0.3)$\times$10$^{12}$ cm$^{-2}$ for T$_{rot}$=5\,K and 
(9.0$\pm$0.9)$\times$10$^{11}$ cm$^{-2}$ 
for T$_{rot}$=7\,K.

For the hydrocarbon anions C$_4$H$^-$, C$_6$H$^-$, and C$_8$H$^-$, and  for their neutral counterparts 
C$_4$H, C$_6$H, and C$_8$H, column densities were derived from all data available in the literature plus 
the QUIJOTE data. For these species, we derived column densities using both the LVG method and the rotation diagram 
method. In Table~\ref{abundances} we give values of their column densities, where the uncertainty corresponds to the values 
derived from these two methods. More details will be given in a separate publication (Ag\'undez et al., in preparation).

Several transitions of C$_7$H are present within the survey. Their intensities are well reproduced with T$_{rot}$=7\,K
and a column density of (6.5$\pm$0.6)$\times$10$^{10}$ cm$^{-2}$. For C$_9$H we derived only a 3$\sigma$ upper 
limit of 3.5$\times$10$^{10}$ cm$^{-2}$ adopting a rotational temperature of 10\,K.

Several hydrocarbons of interest for this work have a zero dipole moment, and hence their abundances were estimated from
their CCH or CN derivatives. We assumed an abundance ratio between the mother
molecule and its cyano derivatives (benzene and naphthalene) of 5--10. In the case of ethynyl derivatives (C$_2$H$_4$, C$_2$H$_6$) 
the adopted abundance ratio is 3--6; however, these estimates are  very uncertain.

\begin{table*}
\caption{Column density of C$_n$H$_m$, C$_n$H$^+$, and C$_n$H$^-$ species in TMC-1}
\label{abundances}
\centering
\begin{tabular}{{lccr}}
\hline
Molecule               &      N       &   Abundance$^a$ & Ref.\\
                       &  (cm$^{-2}$) &                 &      \\
\hline
CH                     & (6-16)$\times$10$^{13}$          & (6-16)$\times$10$^{-09}$          & 1\\
\hline
CCH                    & (6.5$\pm$2.7)$\times$10$^{14}$ & (6.5$\pm$2.7)$\times$10$^{-08}$ & 2\\   
C$_2$H$_4$             & (3-7)$\times$10$^{13}$         & (3-7)$\times$10$^{-09}$         &3a\\
CH$_3$CH$_3$           & (3-6)$\times$10$^{12}$         & (3-6)$\times$10$^{-10}$      &3b\\
\hline
C$_3$H                 & (6.0$\pm$0.6)$\times$10$^{12}$ & (6.0$\pm$0.6)$\times$10$^{-10}$ & 4\\
$c$-C$_3$H             & (1.2$\pm$0.3)$\times$10$^{12}$ & (1.2$\pm$0.3)$\times$10$^{-10}$ & 4\\
C$_3$H$^+$             & (2.4$\pm$0.2)$\times$10$^{10}$ & (2.4$\pm$0.2)$\times$10$^{-12}$ & 5\\
$c-$C$_3$H$_2$         & (1.2$\pm$0.1)$\times$10$^{14}$ & (1.2$\pm$0.1)$\times$10$^{-08}$ & 6\\ 
H$_2$C$_3$             & (1.9$\pm$0.5)$\times$10$^{12}$ & (1.9$\pm$0.5)$\times$10$^{-10}$ & 7\\
CH$_2$CCH              & (8.7$\pm$0.7)$\times$10$^{13}$ & (8.7$\pm$0.7)$\times$10$^{-09}$ & 8\\
CH$_3$CCH              & (1.3$\pm$0.1)$\times$10$^{14}$ & (1.3$\pm$0.1)$\times$10$^{-08}$ & 6\\ 
CH$_2$CHCH$_3$         & (4.0$\pm$0.4)$\times$10$^{13}$ & (4.0$\pm$0.4)$\times$10$^{-09}$ & 9\\
\hline                                                                                        
C$_4$H                 & (9.0$\pm$1.0)$\times$10$^{13}$ & (9.9$\pm$1.0)$\times$10$^{-09}$ & 10\\ 
C$_4$H$^-$             & (2.1$\pm$0.2)$\times$10$^{10}$ & (2.1$\pm$0.2)$\times$10$^{-12}$ & 10\\ 
H$_2$C$_4$             & (3.3$\pm$0.9)$\times$10$^{12}$ & (1.3$\pm$0.1)$\times$10$^{-10}$ & 7\\ 
CH$_2$CHCCH            & (1.2$\pm$0.2)$\times$10$^{13}$ & (1.2$\pm$0.2)$\times$10$^{-09}$ &11\\
\hline
C$_5$H                 & (1.3$\pm$0.3)$\times$10$^{12}$ & (1.3$\pm$0.3)$\times$10$^{-10}$ &12\\
$c$-C$_5$H             & (9.0$\pm$0.9)$\times$10$^{10}$ & (9.0$\pm$0.9)$\times$10$^{-12}$ &12\\
C$_5$H$^+$             & (8.8$\pm$0.5)$\times$10$^{10}$ & (8.8$\pm$0.5)$\times$10$^{-12}$ & 5\\
$c$-C$_3$HCCH          & (3.1$\pm$0.8)$\times$10$^{11}$ & (3.1$\pm$0.8)$\times$10$^{-11}$ &13\\
H$_2$C$_5$             & (1.8$\pm$0.5)$\times$10$^{10}$ & (1.8$\pm$0.5)$\times$10$^{-12}$ & 7\\
CH$_2$CCHCCH           & (1.2$\pm$0.2)$\times$10$^{13}$ & (1.2$\pm$0.2)$\times$10$^{-09}$ &14\\
CH$_3$C$_4$H           & (6.5$\pm$0.3)$\times$10$^{12}$ & (6.5$\pm$0.3)$\times$10$^{-10}$ &14\\
$c$-C$_5$H$_6$         & (1.2$\pm$0.3)$\times$10$^{13}$ & (1.2$\pm$0.3)$\times$10$^{-09}$ &13\\
\hline
C$_6$H                 & (6.0$\pm$1.0)$\times$10$^{12}$ & (6.0$\pm$1.0)$\times$10$^{-10}$ &10\\
C$_6$H$^-$             & (1.6$\pm$0.3)$\times$10$^{11}$ & (1.6$\pm$0.3)$\times$10$^{-11}$ &10\\
H$_2$C$_6$             & (8.0$\pm$2.4)$\times$10$^{10}$ & (8.0$\pm$2.4)$\times$10$^{-12}$ & 7\\ 
$o$-C$_6$H$_4$         & (5.0$\pm$1.0)$\times$10$^{11}$ & (5.0$\pm$1.0)$\times$10$^{-11}$ &15\\
$c$-C$_6$H$_6$         & (0.6-1.2)$\times$10$^{13}$     & (0.6-1.2)$\times$10$^{-09}$     &16\\
$c$-C$_5$H$_4$CH$_2$   & $\le$3.5$\times$10$^{12}$      & $\le$3.5$\times$10$^{-10}$      &17\\
\hline
C$_7$H                 & (6.5$\pm$0.6)$\times$10$^{10}$ & (6.5$\pm$0.6)$\times$10$^{-12}$ & 4\\
CH$_2$CCHC$_4$H        & (1.2$\pm$0.2)$\times$10$^{11}$ & (1.2$\pm$0.2)$\times$10$^{-11}$ &18\\
CH$_3$C$_6$H           & (7.0$\pm$0.7)$\times$10$^{11}$ & (7.0$\pm$0.7)$\times$10$^{-11}$ &18\\
1-$c$-C$_5$H$_5$CCH    & (1.4$\pm$0.2)$\times$10$^{12}$ & (1.4$\pm$0.2)$\times$10$^{-10}$ &19\\
2-$c$-C$_5$H$_5$CCH    & (2.0$\pm$0.4)$\times$10$^{12}$ & (2.0$\pm$0.2)$\times$10$^{-10}$ &19\\
\textbf{$c$-C$_5$H$_4$CCH$_2$}  & (2.7$\pm$0.3)$\times$10$^{12}$ & (2.7$\pm$0.3)$\times$10$^{-10}$ &17\\
$c$-C$_6$H$_5$CH$_3$   & $\le$6.0$\times$10$^{12}$      & $\le$6.0$\times$10$^{-10}$      &17\\
\hline
C$_8$H                 & (5.0$\pm$2.1)$\times$10$^{11}$ & (5.0$\pm$2.1)$\times$10$^{-11}$ & 10\\
C$_8$H$^-$             & (2.6$\pm$0.4)$\times$10$^{10}$ & (2.6$\pm$0.4)$\times$10$^{-12}$ & 10\\
$c$-C$_6$H$_5$CHCH$_2$ & $\le$1.0$\times$10$^{14}$      & $\le$1.0$\times$10$^{-08}$      &17\\
\hline
C$_9$H                 &  $\le$3.5$\times$10$^{10}$     & $\le$3.5$\times$10$^{-12}$      & 4\\
$c$-C$_9$H$_8$         & (1.6$\pm$0.3)$\times$10$^{13}$ & (1.6$\pm$0.3)$\times$10$^{-09}$ &13\\
\hline
1-$c$-C$_{10}$H$_8$    & (3.5-7)$\times$10$^{12}$       & (3.5-7)$\times$10$^{-10}$       &20\\   
2-$c$-C$_{10}$H$_8$    & (3.5-7)$\times$10$^{12}$       & (3.5-7)$\times$10$^{-10}$       &20\\   
\end{tabular}
\tablefoot{\\
        \tablefoottext{a}{Assuming a H$_2$ column density of 10$^{22}$ cm$^{-2}$ \citep{Cernicharo1987}.}\\
        \tablefoottext{1}{Range of column densities derived by \citet{Hjalmarson1977} towards several positions in
        Taurus, some of them around TMC-1. The quoted column densities of CH are representative of the Taurus complex.}
        \tablefoottext{2}{\citet{Sakai2010}.} 
        \tablefoottext{3a}{Estimated from the column density of CH$_2$CHCCH obtained by \citet{Cernicharo2021d} and
        assuming an abundance ratio between  C$_2$H$_4$ and CH$_2$CHCCH $\sim$3--6.}
        \tablefoottext{3b}{Estimated from the column density of CH$_3$CH$_2$CCH obtained by \citet{Cernicharo2021d} and
        assuming an abundance ratio between C$_2$H$_6$ and CH$_3$CH$_2$CCH $\sim$3--6.}
        \tablefoottext{4}{See Appendix \ref{abundance_hydrocarbons}}
        \tablefoottext{5}{\citet{Cernicharo2022}.}
        \tablefoottext{6}{\citet{Cabezas2021a}.}
        \tablefoottext{7}{\citet{Cabezas2021c}.}
        \tablefoottext{8}{\citet{Agundez2021}.}
        \tablefoottext{9}{\citet{Marcelino2007}.}
        \tablefoottext{10}{Ag\'undez et al., in preparation.}
        \tablefoottext{11}{\citet{Cernicharo2021d}.}
        \tablefoottext{12}{\citet{Cabezas2022b}.}
        \tablefoottext{13}{\citet{Cernicharo2021c}.}
        \tablefoottext{14}{\citet{Cernicharo2021b}.}
        \tablefoottext{15}{\citet{Cernicharo2021a}.}
        \tablefoottext{16}{Estimated from the column density of C$_6$H$_5$CN obtained by \citet{Cernicharo2021f} and
        assuming an abundance ratio between benzene and benzonitrile $\sim$5--10.}
        \tablefoottext{17}{This work.}
        \tablefoottext{18}{\citet{Fuentetaja2022}.}
        \tablefoottext{19}{\citet{Cernicharo2021f}.}
        \tablefoottext{20}{From the column density of the isomers of cyanonaphthalene derived by \citet{McGuire2021} assuming
        an abundance ratio of 5--10 between naphthalene and its cyano derivatives.}
}
\end{table*}
\normalsize                

\section{Data analysis} \label{data_analysis}

The raw data of the QUIJOTE survey was processed always following  the same procedure for each observing run, which
typically consist of 100--120 hours of observing time on the source. Each FFT spectrometer
was divided into 160 subsections of 400 channels. Ripples with frequencies $\le$5 MHz can be easily removed with the FFT command
of the CLASS program of the GILDAS package. For each sub-band of 400 channels, a comparison of the sum of the raw data (averaged weighting by system temperature and integration time), the data with ripples was removed, and the results of the previous runs was performed to define line windows
around features with intensity above 4$\sigma$, and also to remove
instrumental spurious features,  atmospheric lines, and well-known features arising in the down-conversion chain.
Once this process was achieved, then all raw scans of a given run were processed removing a polynomial baseline.
Taking into account that the lines have a width of 2--3  channels at half intensity,
this baseline removal does not affect the line profiles. Once the baselines for all the scans of each 400-channel sub-band were
removed, the data were averaged, but this time with weights proportional to 1/$\sigma^2$. Because each run consists typically of 1000--2000 individual scans this procedure
 gives little statistical weight to scans that have a greater noise value   than expected without  having to explore
all the individual scans. This rather tedious procedure yields 
flat baselines and  full confidence in the spectral purity. A total of 2560 blocks of 400 channels have been analysed for each observing run
(160$\times$ 8 spectrometers $\times$ two polarisations). Except for the first run for which the information used on the lines appearing in
TMC-1 was that provided by the Nobeyama line survey of TMC-1 \citep{Kaifu2004}, for all the other runs QUIJOTE itself was the source of
information on the spectral features present in each 400-channel sub-band. The final spectrum contains 507017 channels with a width of 38.15 kHz.

\section{Line parameters} \label{line_parameters}

The line parameters in this work were obtained by fitting a Gaussian line
profile to the observed data. A window of $\pm$ 15 \kms\, around the v$_{LSR}$ of the source (5.83 km s$^{-1}$)
was considered for each transition. The derived line parameters for fulvenallene are given in Table \ref{line_par_fva}.
As indicated in Sect.~\ref{results}, a merged fit of the derived frequencies in TMC-1 with the laboratory measurements of
\citet{Sakaizumi1993} and \citet{McCarthy2020} was performed to improve the rotational and
distortion constants. The list of lines, observed and calculated frequencies, and the difference between observed
and calculated values are given in Table \ref{line_fit_dif_fva}. The recommended rotational and distortion constants
are given in Table \ref{line_fit_fva}.

\begin{table*}
\caption{Observed line parameters for fulvanallene ($c$-C$_5$H$_4$CCH$_2$) in TMC-1}
\label{line_par_fva}
\centering
\begin{tabular}{{cccccc}}
\hline
{\textit Transition} & $\nu_{obs}^a$       & $\int$T$_A^*$dv $^b$     & $\Delta$v$^c$   & T$_A^*$& N\\
                     &  (MHz)              & (mK km\,s$^{-1}$)        & (km\,s$^{-1}$)  & (mK)   & \\
\hline
$ 9_{3, 6}- 8_{3, 5}$&31217.308$\pm$0.020&0.24$\pm$0.10&0.51$\pm$0.41&0.45$\pm$0.16& \\
$ 9_{2, 7}- 8_{2, 6}$&31983.358$\pm$0.015&0.43$\pm$0.08&0.75$\pm$0.14&0.53$\pm$0.10& \\
$ 9_{1, 8}- 8_{1, 7}$&31983.739$\pm$0.010&0.78$\pm$0.13&1.16$\pm$0.24&0.63$\pm$0.10& \\
$10_{1,10}- 9_{1, 9}$&32255.854$\pm$0.010&0.36$\pm$0.08&0.73$\pm$0.17&0.46$\pm$0.10& \\
$10_{0,10}- 9_{0, 9}$&32699.259$\pm$0.015&0.27$\pm$0.08&0.82$\pm$0.31&0.31$\pm$0.10& \\
$10_{2, 9}- 9_{2, 8}$&34027.832$\pm$0.015&0.28$\pm$0.06&0.92$\pm$0.24&0.28$\pm$0.09& \\
$10_{3, 8}- 9_{3, 7}$&34539.679$\pm$0.010&0.34$\pm$0.07&0.72$\pm$0.16&0.45$\pm$0.10& \\
$10_{3, 7}- 9_{3, 6}$&34781.417$\pm$0.010&0.51$\pm$0.12&0.71$\pm$0.18&0.67$\pm$0.15&A\\
$10_{1, 9}- 9_{1, 8}$&35389.366$\pm$0.010&0.50$\pm$0.09&0.69$\pm$0.14&0.68$\pm$0.10& \\
$11_{1,11}-10_{1,10}$&35409.019$\pm$0.010&0.59$\pm$0.06&0.97$\pm$0.24&0.58$\pm$0.14&B\\
$10_{2, 8}- 9_{2, 7}$&35675.155$\pm$0.015&0.42$\pm$0.10&1.02$\pm$0.40&0.39$\pm$0.12& \\
$11_{0,11}-10_{0,10}$&35766.192$\pm$0.015&0.24$\pm$0.05&0.62$\pm$0.11&0.36$\pm$0.09& \\
$11_{2,10}-10_{2, 9}$&37349.808$\pm$0.015&0.37$\pm$0.07&0.75$\pm$0.15&0.46$\pm$0.12&A\\
$11_{3, 9}-10_{3, 8}$&38002.363$\pm$0.020&0.10$\pm$0.04&0.41$\pm$0.14&0.23$\pm$0.09& \\
$11_{3, 8}-10_{3, 7}$&38383.903$\pm$0.020&0.35$\pm$0.09&0.55$\pm$0.21&0.60$\pm$0.14&B\\
$12_{1,12}-11_{1,11}$&38550.784$\pm$0.015&0.23$\pm$0.07&0.66$\pm$0.26&0.44$\pm$0.13&B\\
$11_{1,10}-10_{1, 9}$&38739.886$\pm$0.010&0.37$\pm$0.05&0.60$\pm$0.09&0.58$\pm$0.09& \\
$12_{0,12}-11_{0,11}$&38829.515$\pm$0.015&0.29$\pm$0.10&0.79$\pm$0.42&0.35$\pm$0.11& \\
$11_{2, 9}-10_{2, 8}$&39354.949$\pm$0.030&0.57$\pm$0.05&0.65$\pm$0.08&0.83$\pm$0.13&C\\
$12_{2,11}-11_{2,10}$&40650.743$\pm$0.020&0.42$\pm$0.49&1.09$\pm$0.80&0.36$\pm$0.13&D\\
$12_{3,10}-11_{3, 9}$&41458.231$\pm$0.000&                          &               &$\le$0.63    &A,E\\
$13_{1,13}-12_{1,12}$&41682.937$\pm$0.000&                          &               &$\le$0.50    &F\\
$13_{0,13}-12_{0,12}$&41894.767$\pm$0.000&                          &               &$\le$0.75    &B\\
$12_{3, 9}-11_{3, 8}$&42028.838$\pm$0.020&0.24$\pm$0.11&0.41$\pm$0.19&0.54$\pm$0.14&F\\
$12_{1,11}-11_{1,10}$&42029.496$\pm$0.020&0.54$\pm$0.08&0.61$\pm$0.11&0.82$\pm$0.13& \\ 
$12_{2,10}-11_{2, 9}$&43009.536$\pm$0.015&0.40$\pm$0.12&0.49$\pm$0.19&0.76$\pm$0.20&A\\
$13_{2,12}-12_{2,11}$&43930.042$\pm$0.000&                          &               &$\le$0.54    & \\
$14_{1,14}-13_{1,13}$&44806.950$\pm$0.010&0.54$\pm$0.11&0.63$\pm$0.12&0.81$\pm$0.18& \\
$13_{3,11}-12_{3,10}$&44904.123$\pm$0.030&0.71$\pm$0.13&1.26$\pm$0.24&0.53$\pm$0.16&G\\
$14_{0,14}-13_{0,13}$&44964.612$\pm$0.015&0.20$\pm$0.07&0.42$\pm$0.15&0.45$\pm$0.16& \\
$13_{1,12}-12_{1,11}$&45255.582$\pm$0.015&1.00$\pm$0.18&1.24$\pm$0.26&0.76$\pm$0.17&B\\
$13_{3,10}-12_{3, 9}$&45716.707$\pm$0.000&                          &               &$\le$0.60    & \\
$13_{2,11}-12_{2,10}$&46628.461$\pm$0.000&                          &               &$\le$0.75    &H,B\\ 
$14_{2,13}-13_{2,12}$&47187.614$\pm$0.000&                          &               &$\le$0.54    & \\
$15_{1,15}-14_{1,14}$&47924.500$\pm$0.020&0.47$\pm$0.09&0.60$\pm$0.13&0.73$\pm$0.21& \\
$15_{0,15}-14_{0,14}$&48039.847$\pm$0.020&0.58$\pm$0.12&0.94$\pm$0.16&0.58$\pm$0.22& \\
$14_{3,12}-13_{3,11}$&48336.806$\pm$0.000&                          &               &$\le$0.90      \\
$14_{1,13}-13_{1,12}$&48420.601$\pm$0.020&0.46$\pm$0.17&0.83$\pm$0.25&0.52$\pm$0.28&I,B\\
$14_{3,11}-13_{3,10}$&49443.118$\pm$0.000&                          &               &$\le$1.0     & \\
\hline
\end{tabular}
\tablefoot{\\
        \tablefoottext{a}{Observed frequencies adopting a v$_{LSR}$ of 5.83 km s$^{-1}$ for \mbox{TMC-1}.}\\
        \tablefoottext{b}{Integrated line intensity in mK km\,s$^{-1}$.}\\
        \tablefoottext{c}{Linewidth at half intensity derived by fitting a Gaussian line profile to the observed
     transitions (in km\,s$^{-1}$).}\\
\tablefoottext{A}{Frequency switching data with a 10 MHz throw only. Negative feature present in the data with an 8 MHz throw.}\\
\tablefoottext{B}{Frequency switching data with an 8 MHz throw only. Negative feature present in the data with a 10 MHz throw.}\\
\tablefoottext{C}{Blended exactly at the same frequency with a line of H$_2$C$_4$N that contributes up to 60\% of the observed feature.
  Assuming the remaining intensity is produced by this transition of fulvenallene, then the frequency of the
  transition may be 39354.949$\pm$0.030 MHz.}\\
\tablefoottext{D}{Blended with another feature. Fit still possible.}\\
\tablefoottext{E}{Upper limits correspond to three times the rms of the data. In these cases, frequencies correspond to
  those calculated with the recommended constants of Table \ref{line_fit_fva}.}\\
\tablefoottext{F}{Negative feature in both frequency switching data sets.}\\
\tablefoottext{G}{Line too broad. Probably blended with another feature.}\\
\tablefoottext{H}{Blended with a strong U feature. Unreliable fit.}\\
\tablefoottext{I}{Marginal detection.}\\
}
\end{table*}
\normalsize

\begin{longtable}{crrrrrc}
\caption{Observed and calculated frequencies of $c$-C$_5$H$_4$CCH$_2$} \label{line_fit_dif_fva}\\
\hline
Transition & \multicolumn{1}{c}{$\nu_{obs}^a$} & \multicolumn{1}{c}{$\Delta\nu_{obs}^b$} & \multicolumn{1}{c}{$\nu_{cal}^c$} & \multicolumn{1}{c}{$\Delta\nu_{cal}^d$} & \multicolumn{1}{c}{$\nu_{obs}-\nu_{cal}^e$} & Ref \\
&  \multicolumn{1}{c}{(MHz)}  &  \multicolumn{1}{c}{(MHz)}  &  \multicolumn{1}{c}{(MHz)}  &   \multicolumn{1}{c}{(MHz)}    &  \multicolumn{1}{c}{(MHz)}  &    \\
\hline
\endfirsthead
\caption{continued.}\\
\hline
Transition & \multicolumn{1}{c}{$\nu_{obs}^a$} & \multicolumn{1}{c}{$\Delta\nu_{obs}^b$} & \multicolumn{1}{c}{$\nu_{cal}^c$} & \multicolumn{1}{c}{$\Delta\nu_{cal}^d$} & \multicolumn{1}{c}{$\nu_{obs}-\nu_{cal}^e$} & Ref \\
&  \multicolumn{1}{c}{(MHz)}  &  \multicolumn{1}{c}{(MHz)}  &  \multicolumn{1}{c}{(MHz)}  &   \multicolumn{1}{c}{(MHz)}    &  \multicolumn{1}{c}{(MHz)}  &    \\
\hline
\endhead
\hline
\endfoot
\hline
    2$_{ 1,  2}$- 1$_{ 1,  1}$ &  6533.9975 & 0.0020 &  6533.9997 & 0.0004 & $-$0.0022 & 1 \\
    2$_{ 0,  2}$- 1$_{ 0,  1}$ &  6858.3532 & 0.0020 &  6858.3531 & 0.0002 & $ $0.0001 & 1 \\
    2$_{ 1,  1}$- 1$_{ 1,  0}$ &  7209.1178 & 0.0020 &  7209.1148 & 0.0004 & $ $0.0030 & 1 \\
    3$_{ 1,  3}$- 2$_{ 1,  2}$ &  9792.8921 & 0.0020 &  9792.8930 & 0.0006 & $-$0.0009 & 1 \\
    3$_{ 0,  3}$- 2$_{ 0,  2}$ & 10254.6211 & 0.0020 & 10254.6211 & 0.0004 & $ $0.0000 & 1 \\
    3$_{ 2,  2}$- 2$_{ 2,  1}$ & 10307.2815 & 0.0020 & 10307.2792 & 0.0003 & $ $0.0023 & 1 \\
    3$_{ 2,  1}$- 2$_{ 2,  0}$ & 10360.0099 & 0.0020 & 10360.0094 & 0.0004 & $ $0.0005 & 1 \\
    3$_{ 1,  2}$- 2$_{ 1,  1}$ & 10805.2427 & 0.0020 & 10805.2412 & 0.0006 & $ $0.0015 & 1 \\
    4$_{ 1,  4}$- 3$_{ 1,  3}$ & 13042.5047 & 0.0020 & 13042.5048 & 0.0007 & $-$0.0001 & 1 \\
    4$_{ 0,  4}$- 3$_{ 0,  3}$ & 13612.1720 & 0.0020 & 13612.1735 & 0.0005 & $-$0.0015 & 1 \\
    4$_{ 2,  3}$- 3$_{ 2,  2}$ & 13732.7538 & 0.0020 & 13732.7509 & 0.0004 & $ $0.0029 & 1 \\
    4$_{ 3,  2}$- 3$_{ 3,  1}$ & 13768.4019 & 0.0020 & 13768.4014 & 0.0005 & $ $0.0005 & 1 \\
    4$_{ 3,  1}$- 3$_{ 3,  0}$ & 13770.3390 & 0.0020 & 13770.3417 & 0.0005 & $-$0.0027 & 1 \\
    4$_{ 2,  2}$- 3$_{ 2,  1}$ & 13863.7014 & 0.0020 & 13863.7012 & 0.0006 & $ $0.0002 & 1 \\
    4$_{ 1,  3}$- 3$_{ 1,  2}$ & 14390.7930 & 0.0020 & 14390.7923 & 0.0007 & $ $0.0007 & 1 \\
    5$_{ 1,  5}$- 4$_{ 1,  4}$ & 16280.4664 & 0.0020 & 16280.4654 & 0.0009 & $ $0.0010 & 1 \\
    5$_{ 0,  5}$- 4$_{ 0,  4}$ & 16920.4988 & 0.0020 & 16920.5009 & 0.0007 & $-$0.0021 & 1 \\
    5$_{ 2,  4}$- 4$_{ 2,  3}$ & 17149.4248 & 0.0020 & 17149.4194 & 0.0005 & $ $0.0054 & 1 \\
    5$_{ 3,  3}$- 4$_{ 3,  2}$ & 17220.7569 & 0.0020 & 17220.7543 & 0.0006 & $ $0.0026 & 1 \\
    5$_{ 3,  2}$- 4$_{ 3,  1}$ & 17227.5324 & 0.0020 & 17227.5312 & 0.0006 & $ $0.0012 & 1 \\
    5$_{ 2,  3}$- 4$_{ 2,  2}$ & 17407.7712 & 0.0020 & 17407.7712 & 0.0009 & $ $0.0000 & 1 \\
    5$_{ 1,  4}$- 4$_{ 1,  3}$ & 17961.4705 & 0.0020 & 17961.4710 & 0.0008 & $-$0.0005 & 1 \\
    6$_{ 1,  6}$- 5$_{ 1,  5}$ & 19504.9824 & 0.0020 & 19504.9813 & 0.0010 & $ $0.0011 & 1 \\
    6$_{ 0,  6}$- 5$_{ 0,  5}$ & 20172.9326 & 0.0020 & 20172.9377 & 0.0011 & $-$0.0051 & 1 \\
    6$_{ 2,  5}$- 5$_{ 2,  4}$ & 20555.1800 & 0.0500 & 20555.1215 & 0.0005 & $ $0.0586 & 2 \\
    6$_{ 5,  1}$- 5$_{ 5,  0}$ & 20646.8700 & 0.0500 & 20646.8382 & 0.0020 & $ $0.0318 & 2 \\
    6$_{ 5,  2}$- 5$_{ 5,  1}$ & 20646.8700 & 0.0500 & 20646.8370 & 0.0020 & $ $0.0330 & 2 \\
    6$_{ 3,  4}$- 5$_{ 3,  3}$ & 20678.2900 & 0.0500 & 20678.2760 & 0.0007 & $ $0.0140 & 2 \\
    6$_{ 3,  3}$- 5$_{ 3,  2}$ & 20696.3100 & 0.0500 & 20696.2793 & 0.0008 & $ $0.0307 & 2 \\
    6$_{ 2,  4}$- 5$_{ 2,  3}$ & 20996.5800 & 0.0500 & 20996.6225 & 0.0013 & $-$0.0425 & 2 \\
    6$_{ 1,  5}$- 5$_{ 1,  4}$ & 21512.3113 & 0.0020 & 21512.3128 & 0.0009 & $-$0.0015 & 1 \\
    7$_{ 1,  7}$- 6$_{ 1,  6}$ & 22714.9500 & 0.0500 & 22714.9139 & 0.0011 & $ $0.0361 & 2 \\
    7$_{ 0,  7}$- 6$_{ 0,  6}$ & 23368.6900 & 0.0500 & 23368.6772 & 0.0015 & $ $0.0128 & 2 \\
    7$_{ 2,  6}$- 6$_{ 2,  5}$ & 23947.7600 & 0.0500 & 23947.7488 & 0.0006 & $ $0.0112 & 2 \\
    7$_{ 6,  1}$- 6$_{ 6,  0}$ & 24085.9100 & 0.0500 & 24085.9553 & 0.0034 & $-$0.0453 & 2 \\
    7$_{ 6,  2}$- 6$_{ 6,  1}$ & 24085.9100 & 0.0500 & 24085.9553 & 0.0034 & $-$0.0453 & 2 \\
    7$_{ 5,  3}$- 6$_{ 5,  2}$ & 24096.3600 & 0.0500 & 24096.3213 & 0.0022 & $ $0.0388 & 2 \\
    7$_{ 5,  2}$- 6$_{ 5,  1}$ & 24096.3600 & 0.0500 & 24096.3279 & 0.0022 & $ $0.0321 & 2 \\
    7$_{ 3,  5}$- 6$_{ 3,  4}$ & 24140.5000 & 0.0500 & 24140.4484 & 0.0009 & $ $0.0516 & 2 \\
    7$_{ 3,  4}$- 6$_{ 3,  3}$ & 24180.7300 & 0.0500 & 24180.7009 & 0.0010 & $ $0.0291 & 2 \\
    7$_{ 2,  5}$- 6$_{ 2,  4}$ & 24628.6800 & 0.0500 & 24628.7239 & 0.0019 & $-$0.0439 & 2 \\
    7$_{ 1,  6}$- 6$_{ 1,  5}$ & 25037.5386 & 0.0020 & 25037.5424 & 0.0010 & $-$0.0038 & 1 \\
    8$_{ 1,  8}$- 7$_{ 1,  7}$ & 25909.8500 & 0.0500 & 25909.8088 & 0.0013 & $ $0.0412 & 2 \\
    8$_{ 0,  8}$- 7$_{ 0,  7}$ & 26513.6500 & 0.0500 & 26513.6530 & 0.0018 & $-$0.0030 & 2 \\
    8$_{ 2,  7}$- 7$_{ 2,  6}$ & 27325.3300 & 0.0500 & 27325.2866 & 0.0009 & $ $0.0434 & 2 \\
    8$_{ 7,  2}$- 7$_{ 7,  1}$ & 27525.0300 & 0.0500 & 27524.9845 & 0.0054 & $ $0.0455 & 2 \\
    8$_{ 7,  1}$- 7$_{ 7,  0}$ & 27525.0300 & 0.0500 & 27524.9845 & 0.0054 & $ $0.0455 & 2 \\
    8$_{ 6,  2}$- 7$_{ 6,  1}$ & 27534.3500 & 0.0500 & 27534.3463 & 0.0038 & $ $0.0037 & 2 \\
    8$_{ 6,  3}$- 7$_{ 6,  2}$ & 27534.3500 & 0.0500 & 27534.3461 & 0.0038 & $ $0.0039 & 2 \\
    8$_{ 5,  3}$- 7$_{ 5,  2}$ & 27549.6600 & 0.0500 & 27549.6856 & 0.0025 & $-$0.0256 & 2 \\
    8$_{ 5,  4}$- 7$_{ 5,  3}$ & 27549.6600 & 0.0500 & 27549.6590 & 0.0025 & $ $0.0010 & 2 \\
    8$_{ 3,  6}$- 7$_{ 3,  5}$ & 27606.0400 & 0.0500 & 27606.0689 & 0.0012 & $-$0.0289 & 2 \\
    8$_{ 2,  6}$- 7$_{ 2,  5}$ & 28295.7200 & 0.0500 & 28295.6725 & 0.0026 & $ $0.0475 & 2 \\
    8$_{ 1,  7}$- 7$_{ 1,  6}$ & 28530.5400 & 0.0500 & 28530.5023 & 0.0014 & $ $0.0377 & 2 \\
    9$_{ 1,  9}$- 8$_{ 1,  8}$ & 29089.9000 & 0.0500 & 29089.8668 & 0.0016 & $ $0.0332 & 2 \\
    9$_{ 0,  9}$- 8$_{ 0,  8}$ & 29619.2800 & 0.0500 & 29619.2698 & 0.0022 & $ $0.0102 & 2 \\
    9$_{ 2,  8}$- 8$_{ 2,  7}$ & 30685.8700 & 0.0500 & 30685.8656 & 0.0012 & $ $0.0044 & 2 \\
    9$_{ 8,  1}$- 8$_{ 8,  0}$ & 30963.7400 & 0.0500 & 30963.8251 & 0.0080 & $-$0.0851 & 2 \\
    9$_{ 8,  2}$- 8$_{ 8,  1}$ & 30963.7400 & 0.0500 & 30963.8251 & 0.0080 & $-$0.0851 & 2 \\
    9$_{ 7,  3}$- 8$_{ 7,  2}$ & 30972.6200 & 0.0500 & 30972.6010 & 0.0060 & $ $0.0100 & 2 \\
    9$_{ 7,  2}$- 8$_{ 7,  1}$ & 30972.6200 & 0.0500 & 30972.6010 & 0.0060 & $ $0.0100 & 2 \\
    9$_{ 6,  4}$- 8$_{ 6,  3}$ & 30985.7800 & 0.0500 & 30985.7573 & 0.0043 & $ $0.0227 & 2 \\
    9$_{ 6,  3}$- 8$_{ 6,  2}$ & 30985.7800 & 0.0500 & 30985.7581 & 0.0043 & $ $0.0219 & 2 \\
    9$_{ 3,  7}$- 8$_{ 3,  6}$ & 31073.2100 & 0.0500 & 31073.2809 & 0.0016 & $-$0.0709 & 2 \\
    9$_{ 3,  6}$- 8$_{ 3,  5}$ & 31217.3080 & 0.0200 & 31217.3077 & 0.0022 & $ $0.0003 & 3 \\
    9$_{ 3,  6}$- 8$_{ 3,  5}$ & 31217.2000 & 0.0500 & 31217.3077 & 0.0022 & $-$0.1077 & 2 \\
    9$_{ 2,  7}$- 8$_{ 2,  6}$ & 31983.3580 & 0.0150 & 31983.3859 & 0.0033 & $-$0.0279 & 3 \\
    9$_{ 1,  8}$- 8$_{ 1,  7}$ & 31983.7390 & 0.0100 & 31983.7384 & 0.0019 & $ $0.0007 & 3 \\
   10$_{ 1, 10}$- 9$_{ 1,  9}$ & 32255.8540 & 0.0100 & 32255.8600 & 0.0020 & $-$0.0060 & 3 \\
   10$_{ 0, 10}$- 9$_{ 0,  9}$ & 32699.2590 & 0.0150 & 32699.2546 & 0.0024 & $ $0.0044 & 3 \\
   10$_{ 0, 10}$- 9$_{ 0,  9}$ & 32699.2400 & 0.0500 & 32699.2546 & 0.0024 & $-$0.0146 & 2 \\
   10$_{ 2,  9}$- 9$_{ 2,  8}$ & 34027.8320 & 0.0150 & 34027.8258 & 0.0016 & $ $0.0063 & 3 \\
   10$_{ 2,  9}$- 9$_{ 2,  8}$ & 34027.8100 & 0.0500 & 34027.8258 & 0.0016 & $-$0.0158 & 2 \\
   10$_{ 9,  1}$- 9$_{ 9,  0}$ & 34402.4400 & 0.0500 & 34402.4085 & 0.0114 & $ $0.0315 & 2 \\
   10$_{ 9,  2}$- 9$_{ 9,  1}$ & 34402.4400 & 0.0500 & 34402.4085 & 0.0114 & $ $0.0315 & 2 \\
   10$_{ 8,  3}$- 9$_{ 8,  2}$ & 34410.8900 & 0.0500 & 34410.8676 & 0.0088 & $ $0.0224 & 2 \\
   10$_{ 8,  2}$- 9$_{ 8,  1}$ & 34410.8900 & 0.0500 & 34410.8676 & 0.0088 & $ $0.0224 & 2 \\
   10$_{ 7,  3}$- 9$_{ 7,  2}$ & 34422.6900 & 0.0500 & 34422.7097 & 0.0066 & $-$0.0197 & 2 \\
   10$_{ 7,  4}$- 9$_{ 7,  3}$ & 34422.6900 & 0.0500 & 34422.7097 & 0.0066 & $-$0.0197 & 2 \\
   10$_{ 6,  4}$- 9$_{ 6,  3}$ & 34440.6000 & 0.0500 & 34440.5737 & 0.0048 & $ $0.0263 & 2 \\
   10$_{ 6,  5}$- 9$_{ 6,  4}$ & 34440.6000 & 0.0500 & 34440.5709 & 0.0048 & $ $0.0292 & 2 \\
   10$_{ 4,  7}$- 9$_{ 4,  6}$ & 34520.9400 & 0.0500 & 34520.9831 & 0.0026 & $-$0.0431 & 2 \\
   10$_{ 4,  6}$- 9$_{ 4,  5}$ & 34532.0500 & 0.0500 & 34532.0806 & 0.0026 & $-$0.0306 & 2 \\
   10$_{ 3,  8}$- 9$_{ 3,  7}$ & 34539.6790 & 0.0100 & 34539.6611 & 0.0023 & $ $0.0179 & 3 \\
   10$_{ 3,  8}$- 9$_{ 3,  7}$ & 34539.6900 & 0.0500 & 34539.6611 & 0.0023 & $ $0.0289 & 2 \\
   10$_{ 3,  7}$- 9$_{ 3,  6}$ & 34781.4170 & 0.0100 & 34781.4410 & 0.0033 & $-$0.0240 & 3 \\
   10$_{ 3,  7}$- 9$_{ 3,  6}$ & 34781.4500 & 0.0500 & 34781.4410 & 0.0033 & $ $0.0090 & 2 \\
   10$_{ 1,  9}$- 9$_{ 1,  8}$ & 35389.3660 & 0.0100 & 35389.3691 & 0.0026 & $-$0.0031 & 3 \\
   10$_{ 1,  9}$- 9$_{ 1,  8}$ & 35389.4100 & 0.0500 & 35389.3691 & 0.0026 & $ $0.0409 & 2 \\
   11$_{ 1, 11}$-10$_{ 1, 10}$ & 35409.0190 & 0.0100 & 35409.0038 & 0.0025 & $ $0.0152 & 3 \\
   11$_{ 1, 11}$-10$_{ 1, 10}$ & 35409.0300 & 0.0500 & 35409.0038 & 0.0025 & $ $0.0262 & 2 \\
   10$_{ 2,  8}$- 9$_{ 2,  7}$ & 35675.1550 & 0.0150 & 35675.1534 & 0.0042 & $ $0.0016 & 3 \\
   10$_{ 2,  8}$- 9$_{ 2,  7}$ & 35675.1200 & 0.0500 & 35675.1534 & 0.0042 & $-$0.0334 & 2 \\
   11$_{ 0, 11}$-10$_{ 0, 10}$ & 35766.1920 & 0.0150 & 35766.2148 & 0.0028 & $-$0.0228 & 3 \\
   11$_{ 0, 11}$-10$_{ 0, 10}$ & 35766.1700 & 0.0500 & 35766.2148 & 0.0028 & $-$0.0448 & 2 \\
   11$_{ 2, 10}$-10$_{ 2,  9}$ & 37349.8080 & 0.0150 & 37349.7879 & 0.0022 & $ $0.0201 & 3 \\
   11$_{ 2, 10}$-10$_{ 2,  9}$ & 37349.7700 & 0.0500 & 37349.7879 & 0.0022 & $-$0.0179 & 2 \\
   11$_{10,  1}$-10$_{10,  0}$ & 37840.6200 & 0.0500 & 37840.6800 & 0.0156 & $-$0.0600 & 2 \\
   11$_{10,  2}$-10$_{10,  1}$ & 37840.6200 & 0.0500 & 37840.6800 & 0.0156 & $-$0.0600 & 2 \\
   11$_{ 9,  3}$-10$_{ 9,  2}$ & 37848.9500 & 0.0500 & 37848.9767 & 0.0124 & $-$0.0267 & 2 \\
   11$_{ 9,  2}$-10$_{ 8,  1}$ & 37848.9500 & 0.0500 & 37848.9767 & 0.0124 & $-$0.0267 & 2 \\
   11$_{ 8,  3}$-10$_{ 8,  2}$ & 37860.0100 & 0.0500 & 37860.0017 & 0.0097 & $ $0.0083 & 2 \\
   11$_{ 8,  4}$-10$_{ 8,  3}$ & 37860.0100 & 0.0500 & 37860.0017 & 0.0097 & $ $0.0083 & 2 \\
   11$_{ 7,  4}$-10$_{ 7,  3}$ & 37875.5500 & 0.0500 & 37875.5615 & 0.0073 & $-$0.0115 & 2 \\
   11$_{ 7,  5}$-10$_{ 7,  4}$ & 37875.5500 & 0.0500 & 37875.5614 & 0.0073 & $-$0.0114 & 2 \\
   11$_{ 6,  5}$-10$_{ 6,  4}$ & 37899.1900 & 0.0500 & 37899.1807 & 0.0054 & $ $0.0093 & 2 \\
   11$_{ 6,  6}$-10$_{ 6,  5}$ & 37899.1900 & 0.0500 & 37899.1717 & 0.0054 & $ $0.0183 & 2 \\
   11$_{ 3,  9}$-10$_{ 3,  8}$ & 38002.3630 & 0.0200 & 38002.3561 & 0.0031 & $ $0.0069 & 3 \\
   11$_{ 4,  7}$-10$_{ 4,  6}$ & 38024.7700 & 0.0500 & 38024.8792 & 0.0035 & $-$0.1092 & 2 \\
   11$_{ 3,  8}$-10$_{ 3,  7}$ & 38383.9030 & 0.0200 & 38383.9494 & 0.0048 & $-$0.0464 & 3 \\
   11$_{ 3,  8}$-10$_{ 3,  7}$ & 38383.9600 & 0.0500 & 38383.9494 & 0.0048 & $ $0.0106 & 2 \\
   12$_{ 1, 12}$-11$_{ 1, 11}$ & 38550.7840 & 0.0150 & 38550.8045 & 0.0034 & $-$0.0205 & 3 \\
   11$_{ 1, 10}$-10$_{ 1,  9}$ & 38739.8860 & 0.0100 & 38739.8829 & 0.0035 & $ $0.0031 & 3 \\
   11$_{ 1, 10}$-10$_{ 1,  9}$ & 38739.8600 & 0.0500 & 38739.8829 & 0.0035 & $-$0.0229 & 2 \\
   12$_{ 0, 12}$-11$_{ 0, 11}$ & 38829.5150 & 0.0150 & 38829.4924 & 0.0035 & $ $0.0226 & 3 \\
   11$_{ 2,  9}$-10$_{ 2,  8}$ & 39354.9490 & 0.0300 & 39354.9605 & 0.0053 & $-$0.0115 & 3 \\
   12$_{ 2, 11}$-11$_{ 2, 10}$ & 40650.7430 & 0.0200 & 40650.7284 & 0.0028 & $ $0.0146 & 3 \\
   12$_{ 3,  9}$-11$_{ 3,  8}$ & 42028.8380 & 0.0200 & 42028.8560 & 0.0067 & $-$0.0180 & 3 \\
   12$_{ 1, 11}$-11$_{ 1, 10}$ & 42029.4960 & 0.0200 & 42029.4421 & 0.0044 & $ $0.0539 & 3 \\
   12$_{ 2, 10}$-11$_{ 2,  9}$ & 43009.5360 & 0.0150 & 43009.5066 & 0.0065 & $ $0.0294 & 3 \\
   14$_{ 1, 14}$-13$_{ 1, 13}$ & 44806.9500 & 0.0100 & 44806.9533 & 0.0059 & $-$0.0033 & 3 \\
   13$_{ 3, 11}$-12$_{ 3, 10}$ & 44904.1230 & 0.0300 & 44904.1187 & 0.0052 & $ $0.0043 & 3 \\
   14$_{ 0, 14}$-13$_{ 0, 13}$ & 44964.6120 & 0.0150 & 44964.6223 & 0.0058 & $-$0.0103 & 3 \\
   13$_{ 1, 12}$-12$_{ 1, 11}$ & 45255.5820 & 0.0150 & 45255.5507 & 0.0053 & $ $0.0313 & 3 \\
   15$_{ 1, 15}$-14$_{ 1, 14}$ & 47924.5000 & 0.0200 & 47924.4911 & 0.0076 & $ $0.0089 & 3 \\
   15$_{ 0, 15}$-14$_{ 0, 14}$ & 48039.8470 & 0.0200 & 48039.8698 & 0.0075 & $-$0.0228 & 3 \\
   14$_{ 1, 13}$-13$_{ 1, 12}$ & 48420.6010 & 0.0200 & 48420.5784 & 0.0063 & $ $0.0227 & 3 \\                 
\hline
\end{longtable}
\tablefoot{\\
        \tablefoottext{a}{Observed frequencies.}\\
        \tablefoottext{b}{Uncertainty of the observed frequeciencies (MHz).}\\
        \tablefoottext{c}{Calculated frequencies resulting from the fit (MHz).}\\
        \tablefoottext{d}{Predicted uncertainty of the calculated frequencies (MHz).}\\
        \tablefoottext{e}{Observed minus calculated frequencies (MHz).}\\
\tablefoottext{1}{Laboratoraty data from \citet{McCarthy2020}.}\\
\tablefoottext{2}{Laboratory data from \citet{Sakaizumi1993}}\\
\tablefoottext{3}{This work.}\\
}   
                   
\end{appendix}

\end{document}